\documentclass[final,5p,times,twocolumn]{elsarticle}

\usepackage[utf8]{inputenc}
\usepackage{amsmath,amssymb,amsfonts}
\usepackage{bm}
\usepackage{graphicx}
\usepackage{xcolor}
\usepackage{booktabs}
\usepackage[colorlinks=true,breaklinks=true,linkcolor=blue,citecolor=purple,urlcolor=teal]{hyperref}
\usepackage{url}

\journal{Physics Letters B}

\biboptions{comma,sort&compress}

\begin{document}

\begin{frontmatter}

\title{Channel couplings redirect absorbed flux from peripheral loss to fusion in weakly bound nuclear reactions}

\author[tongji]{Hao Liu}
\author[tongji]{Jin Lei\corref{cor1}}
\cortext[cor1]{Corresponding author}
\ead{jinl@tongji.edu.cn}
\author[tongji]{Zhongzhou Ren}
\address[tongji]{School of Physics Science and Engineering, Tongji University, Shanghai 200092, China}

\begin{abstract}
In reactions of weakly bound nuclei, the absorption cross section mixes two physically distinct contributions: inner capture associated with compound-nucleus formation, and peripheral losses from breakup, transfer, and other direct reactions. Within a framework that combines an ingoing-wave boundary condition (IWBC) at an inner radius with a complex potential in the external region, we derive the exact flux identity $\sigma_{\rm abs}=\sigma_{\rm fusion}+\sigma_W$ from the radial continuity equation. The resulting partition is exact within the adopted CC/CDCC model space and provides a practical diagnostic of where absorbed flux is removed. Applied to $^6$Li+$^{209}$Bi, the analysis reveals that channel couplings qualitatively reorganize the absorbed flux: the dominant absorption mechanism shifts from peripheral loss at sub-barrier energies to inner capture above the barrier, whereas the single-channel baseline remains peripheral-loss dominated throughout. The resulting IWBC-defined inner-capture cross section tracks the measured complete-fusion excitation function with only a modest dependence on the chosen boundary radius. Together with the exact identity $\sigma_{\rm abs}=\sigma_{\rm fusion}+\sigma_W$, this agreement supports interpreting the peripheral term $\sigma_W$ as a major spatial contributor to the well-known CF suppression in weakly bound systems.
\end{abstract}


\end{frontmatter}

\section{Introduction}
\label{sec:intro}

Heavy-ion fusion near the Coulomb barrier is governed by quantum tunneling~\cite{Balantekin98,Hagino2012} and is strongly modified by couplings to intrinsic excitations and reaction channels. For weakly bound projectiles such as $^6$Li and $^7$Li, breakup, transfer, and other direct-reaction channels compete vigorously with fusion~\cite{Canto06,Canto15,RMP_Back14}, and the measured complete-fusion (CF) cross sections are systematically suppressed relative to single-barrier predictions~\cite{Dasgupta04,Diaz-prl07,Jin19}. Understanding this suppression requires distinguishing, within the total flux removed from the elastic channel, between the fraction that genuinely corresponds to inner capture and the fraction lost to peripheral processes. In standard coupled-channels (CC) and continuum-discretized coupled-channels (CDCC) calculations~\cite{Austern87,Hagino_fusion00,PhysRevC.75.044601,Hashimoto2009}, however, different absorptive contributions are typically intertwined through a single imaginary potential, obscuring the dynamical meaning of the lost flux. Not all absorption is fusion; the critical question is \textit{where} the flux is lost.

A physically motivated separation follows from the spatial structure of heavy-ion reactions.
Fusion is associated with deep penetration into the strongly overlapping inner region~\cite{Udagawa85,PhysRevC.98.044617,ADAMIAN1997241}, whereas direct processes are predominantly peripheral~\cite{RMP_Back14,PhysRevC.108.024606}.
This motivates the standard use of an ingoing-wave boundary condition (IWBC) at an inner radius $r_a$ to define fusion through the inward flux~\cite{Hagino2012,ccfull,EISEN1972219}, while non-fusion loss is represented by a complex interaction in the external region.
Several earlier studies have addressed related aspects of this question: Satchler~\cite{Satchler85} related imaginary potentials to the total reaction cross section via the optical theorem; Udagawa et al.~\cite{Udagawa85} separated direct-reaction and compound-nucleus contributions within a local-potential model; and the generalized optical theorem has been applied in CDCC to extract elastic-breakup components from the total absorption~\cite{COTANCH201048,liu2026}. What has remained absent, however, is a compact and exact spatial partition of the absorbed flux that holds in the presence of channel couplings, where off-diagonal coherence between channels renders the decomposition non-trivial.

A related approach was proposed by Hashimoto \textit{et al.}~\cite{Hashimoto2009}, who considered the separation of complete- and incomplete-fusion by decomposing the relevant contributions in coordinate space according to the ranges of the imaginary absorptive potentials. In contrast, the present study uses the IWBC as the mechanism defining fusion through inner capture, and focuses on flux decomposition in the IWBC$+$external-$W$ framework.

In this Letter, we derive such a partition from the radial continuity equation and show that $\sigma_{\rm abs}=\sigma_{\rm fusion}+\sigma_W$ holds exactly within the IWBC$+$external-$W$ framework for both single-channel and multichannel systems.
Applied to $^6$Li+$^{209}$Bi, the analysis reveals that channel couplings qualitatively redistribute the absorbed flux: the dominant absorption mechanism shifts from peripheral loss to inner capture as the energy increases, with the two fractions crossing near the Coulomb barrier. This crossover, absent in the single-channel baseline studied here, provides a spatial perspective on both the near-barrier fusion enhancement and the above-barrier CF suppression.

\section{Formalism}
\label{sec:formalism}

We consider a coupled-channel system in which the radial configuration space is divided at an inner radius $r_a$.
The region $r<r_a$ represents the strongly overlapping fusion domain and is described by an IWBC; the external region $r>r_a$ is governed by a complex coupling interaction $U_{cc'}(r)=V_{cc'}(r)-iW_{cc'}(r)$, in which the diagonal entry $U_{cc}$ contains the optical potential of channel $c$ and the off-diagonal entries $U_{cc'}$ ($c'\neq c$) generate the channel couplings. We take $V$ and $W$ to be Hermitian in channel space, $V_{cc'}^*=V_{c'c}$ and $W_{cc'}^*=W_{c'c}$, with $W$ positive semidefinite as an operator for physical absorption.
A channel index $c$ collects the orbital angular momentum $l_c$ of the relative motion together with the internal quantum numbers of the projectile--target system; the entrance channel is labeled by $c=0$, and all relations hold independently for each total angular momentum $J$ (the $J$ label is suppressed for brevity). Unless stated otherwise, sums over $c$ refer to channels that are open in the asymptotic region and on which the IWBC is imposed at $r_a$.

The coupled radial equations in the external region read
\begin{equation}
u_c''(r)+\!\left[k_c^2-\frac{l_c(l_c\!+\!1)}{r^2}\right]\!u_c
-\frac{2\mu}{\hbar^2}\sum_{c'}U_{cc'}u_{c'}=0,
\label{eq:cc}
\end{equation}
where $\mu$ is the reduced mass and $k_c=\sqrt{2\mu(E-\epsilon_c)}/\hbar$ the channel wave number.
Multiplying by $u_c^*$, subtracting the complex conjugate, summing over $c$, and integrating from $r_a$ to infinity yields
\begin{equation}
\Phi_\infty-\Phi_a
= -\frac{4i\mu}{\hbar^2}\sum_{cc'}\int_{r_a}^{\infty}W_{cc'}(r)\,u_c^*(r)\,u_{c'}(r)\,dr,
\label{eq:continuity}
\end{equation}
with $\Phi(r)\equiv\sum_c(u_c^*u_c'-u_c u_c^{*\prime})$.

The solution $u_c(r)$ is subject to two boundary conditions. At large distances it takes the asymptotic form
\begin{equation}
u_c(r)\!\xrightarrow{r\to\infty}\!\frac{i}{2}\!\left[H_{l_c}^{(-)}(k_c r)\,\delta_{c0}-\sqrt{\frac{k_0}{k_c}}\,S_{c0}\,H_{l_c}^{(+)}(k_c r)\right]\!,
\end{equation}
while the IWBC at $r_a$ imposes a purely ingoing wave in every open channel~\cite{EISEN1972219,ccfull}:
\begin{equation}
u_{c}(r)\xrightarrow[r\to r_a]{}\frac{i}{2}\sqrt{\frac{k_0}{K_c(r)}}\,\mathcal{T}_{c0}\exp\!\left[-i\int_{r_a}^{r}K_c(r')\,dr'\right]\!,
\label{eq:iwbc_multi}
\end{equation}
where $K_c(r)\equiv\sqrt{(2\mu/\hbar^2)[E-\epsilon_c-U_{cc}(r)]-l_c(l_c+1)/r^2}$~\cite{Knoll1976,1957WKB} is the local channel wave number consistent with the radial equation~(\ref{eq:cc}), reducing to the asymptotic $k_c$ as $r\to\infty$, and $\mathcal{T}_{c0}$ is the transmission amplitude from the entrance channel into channel~$c$. Equation~(\ref{eq:iwbc_multi}) is the channel-by-channel WKB form of the IWBC; multichannel calculations typically implement it in the eigenchannel basis at $r_a$~\cite{ccfull}, and the partition derived below is exact with respect to the adopted prescription. In the antisymmetric combination $u_c^*u_c'-u_cu_c'^*$ entering $\Phi$, the real prefactor $K_c(r)^{-1/2}$ cancels identically as an algebraic property of the WKB form, leaving the contribution $-2iK_c(r)|u_c(r)|^2$. Evaluating $\Phi$ at both limits using the Wronskian of the Coulomb--Hankel functions at $r\to\infty$ then gives
\begin{equation}
\begin{aligned}
\Phi_\infty&=-\tfrac{ik_0}{2}\bigl(1-\textstyle\sum_c|S_{c0}|^2\bigr)
\equiv -\tfrac{ik_0}{2}\,P_{\rm abs},\\
\Phi_a&=-\tfrac{ik_0}{2}\textstyle\sum_c|\mathcal{T}_{c0}|^2
\equiv -\tfrac{ik_0}{2}\,P_{\rm inner}.
\end{aligned}
\end{equation}
Substituting into Eq.~(\ref{eq:continuity}) and defining
\begin{equation}
P_W = \frac{8\mu}{\hbar^2 k_0}\sum_{cc'}\int_{r_a}^{\infty}W_{cc'}(r)\,u_c^*(r)\,u_{c'}(r)\,dr,
\end{equation}
one obtains $P_{\rm abs}=P_{\rm inner}+P_W$.
Restoring the sum over $J$ with the standard weight $\pi(2J+1)/k_0^2$ yields
\begin{equation}
\sigma_{\rm abs}=\sigma_{\rm inner}+\sigma_W.
\label{eq:decomp}
\end{equation}
In the present work, we identify the IWBC-defined inner contribution $\sigma_{\rm inner}$ with the fusion cross section, $\sigma_{\rm fusion}$. With this notation, Eq.~(\ref{eq:decomp}) may also be written as
\begin{equation}
\sigma_{\rm abs}=\sigma_{\rm fusion}+\sigma_W.
\label{eq:decomp_fusion}
\end{equation}
Here, the first term represents the flux that crosses the inner boundary and is irreversibly absorbed under the IWBC prescription, whereas the second term represents the flux removed in the external region by the imaginary interaction. In the present work, we interpret $\sigma_{\rm inner}$ as a model proxy for the complete-fusion cross section, $\sigma_{\rm fusion}$, i.e., a measure of the flux captured in the strongly overlapping inner region where compound-nucleus formation is expected to dominate. This identification should be understood as a model definition within the adopted spatial decomposition rather than a strictly exclusive classification of the final reaction outcome. In particular, $\sigma_W$ should not be interpreted as containing only non-fusion processes: for weakly bound systems, it may also include breakup-triggered pathways in which the projectile dissociates outside $r_a$ and the fragments are subsequently captured by the target. Sequential-capture configurations, in which both breakup fragments are absorbed by the target, are physically complete-fusion pathways, so that a part of $\sigma_W$ itself contributes to fusion. This also underlines the importance of choosing a physically reasonable boundary radius $r_a$, so that the inner region can serve as a meaningful proxy for the strongly overlapping capture domain. Therefore, the exactness of Eqs.~(\ref{eq:decomp}) and~(\ref{eq:decomp_fusion}) refers to the flux decomposition itself within the adopted model, whereas its detailed physical interpretation remains dependent on the choice of boundary radius, the IWBC prescription, and the absorptive interaction. The single-channel limit is recovered trivially by restricting $c$ to a single index.

Two structural features of the decomposition deserve emphasis.
First, $P_W$ is not a sum of independent single-channel absorptions: through the off-diagonal terms $W_{cc'}u_c^*u_{c'}$, it carries channel-space coherence.
The external absorption is therefore a collective quantity whose value depends on the full coupled-channel wave function.
For a physical absorptive interaction, the relevant condition is not elementwise positivity of individual $W_{cc'}$ matrix elements, but positive semidefiniteness of the operator $W$: for any channel vector $\bm{\xi}$, $\sum_{cc'}\xi_c^*W_{cc'}\xi_{c'}\ge 0$.
Under this condition, the quadratic form entering the definition of $P_W$ is non-negative after contraction with the coupled-channel wave function, and therefore $P_W\ge 0$ even in the presence of off-diagonal coherence terms.
Second, the physical content of $\sigma_W$ is determined by the adopted model space.
When certain direct-reaction channels are retained explicitly (e.g., elastic breakup), the total reaction cross section becomes $\sigma_R=\sigma_{\rm EBU}+\sigma_{\rm abs}$~\cite{COTANCH201048,liu2026}, and $\sigma_W$ collects only the remaining unresolved peripheral absorption.
For weakly bound systems, $\sigma_W$ may thus contain contributions from incomplete fusion (in which one cluster fragment is captured while the other escapes) as well as other unresolved direct-reaction losses within the adopted channel space.
Closed channels may still affect the dynamics through coupling in the interior but do not contribute directly to the asymptotic flux balance.

\section{Application to $^6$Li+$^{209}$Bi}
\label{sec:results}

\begin{figure}[t]
\centering
\includegraphics[width=\linewidth]{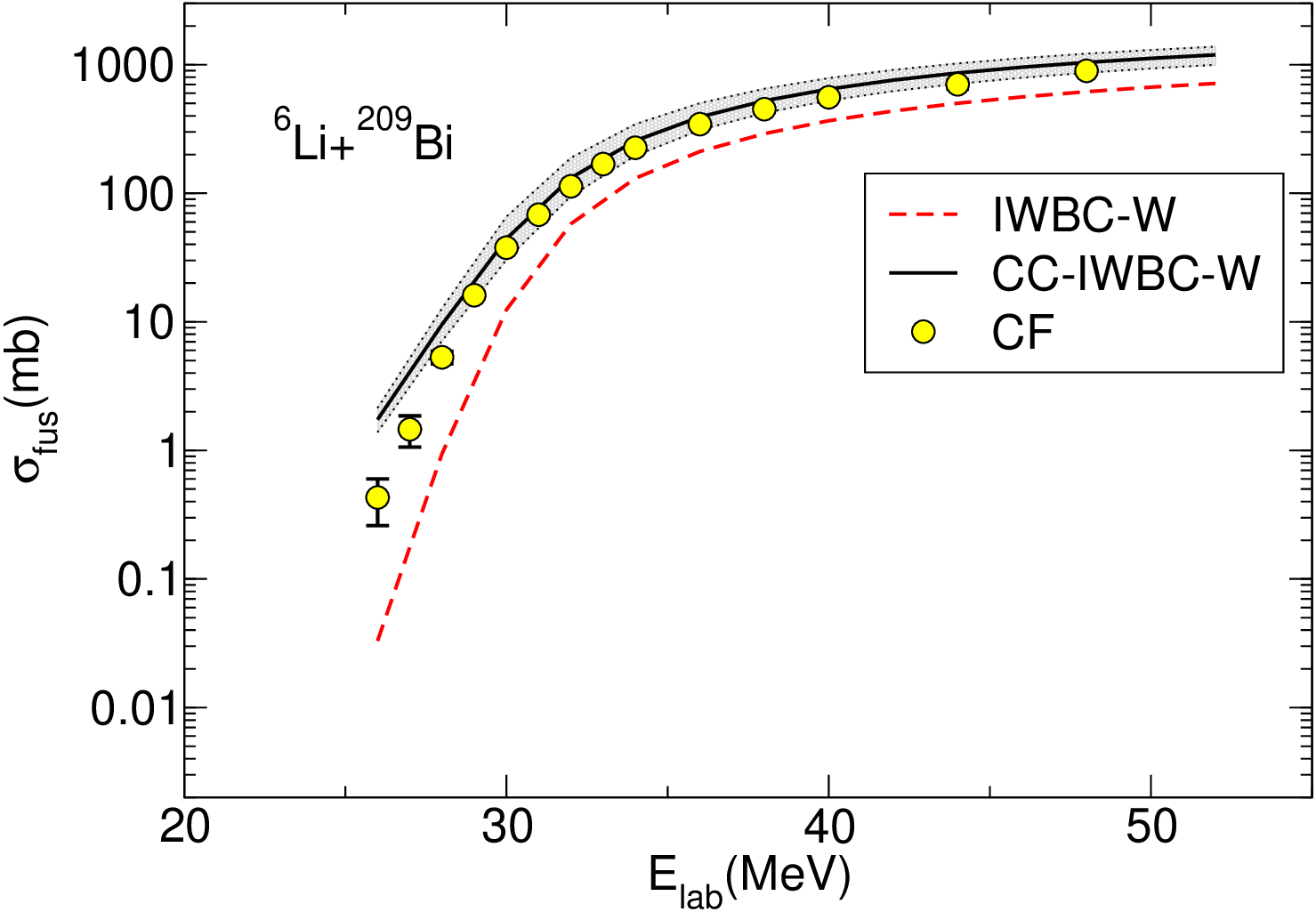}
\caption{IWBC-defined inner-capture cross section $\sigma_{\rm fusion}$ for $^6$Li+$^{209}$Bi, shown for comparison with experimental CF data.
Dashed red line: single-channel; solid black line: coupled-channel ($r_a=10$~fm); gray band: $r_a=9.5$--10.5~fm; yellow circles: experimental CF data~\cite{Dasgupta04}.}
\label{fig:fusion}
\end{figure}

We apply the formalism to $^6$Li+$^{209}$Bi, a benchmark weakly bound system in which breakup and fusion compete strongly near the Coulomb barrier. The IWBC radius is set to $r_a=10$~fm, corresponding to the nuclear touching configuration $R_P+R_T \approx 1.3(A_P^{1/3}+A_T^{1/3})\ \mathrm{fm}$. This is a physically motivated reference rather than an optimized fit parameter; the residual sensitivity to the choice of $r_a$ is quantified by the band in Fig.~\ref{fig:fusion} obtained by varying $r_a$ over $9.5$--$10.5$~fm.

\begin{figure}[t]
\centering
\includegraphics[width=\linewidth]{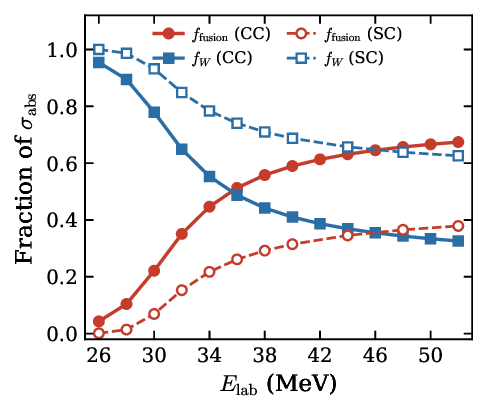}
\caption{Absorption decomposition fractions $f_{\rm fusion}=\sigma_{\rm fusion}/\sigma_{\rm abs}$ (red circles) and $f_W=\sigma_W/\sigma_{\rm abs}$ (blue squares) as functions of $E_{\rm lab}$ for $^6$Li+$^{209}$Bi.
Solid lines and filled symbols: coupled-channel; dashed lines and open symbols: single-channel.}
\label{fig:decomposition}
\end{figure}

As a baseline, we perform a single-channel optical-model calculation with the $^6$Li global optical potential of Ref.~\cite{MASLOV}, fitted to elastic-scattering systematics across target masses $A=12$--208 and providing a physically constrained wave function for the decomposition.
Table~\ref{tab:sc} lists the resulting decomposition at selected energies.
The closure error $\varepsilon/\sigma_{\rm abs}$ remains below $5\times10^{-3}$ over the full energy range, confirming that Eq.~(\ref{eq:decomp}) is satisfied with high numerical precision.
In the single-channel case the external absorption dominates at all energies: the fusion fraction $f_{\rm fusion}=\sigma_{\rm fusion}/\sigma_{\rm abs}$ rises from near zero at 26~MeV to only 0.38 at 52~MeV (dashed lines in Fig.~\ref{fig:decomposition}).

\begin{table}[t]
\caption{Single-channel cross-section decomposition for $^6$Li+$^{209}$Bi with $r_a=10$~fm.
The closure error is listed as a relative quantity $\varepsilon/\sigma_{\rm abs}$.}
\label{tab:sc}
\centering
\begin{tabular}{ccccc}
\toprule
$E_{\rm lab}$ & $\sigma_{\rm abs}$ & $\sigma_{\rm fusion}$ & $\sigma_W$ & $\varepsilon/\sigma_{\rm abs}$ \\
(MeV) & (mb) & (mb) & (mb) & \\
\midrule
26 &    22.4 &    0.03 &    22.4 & $-1.9\!\times\!10^{-3}$ \\
30 &   179.1 &   12.4  &   166.9 & $-1.3\!\times\!10^{-3}$ \\
34 &   601.8 &  130.8  &   471.7 & $-1.3\!\times\!10^{-3}$ \\
38 &   995.9 &  290.9  &   706.8 & $-1.8\!\times\!10^{-3}$ \\
44 &  1450.8 &  501.3  &   953.4 & $-2.7\!\times\!10^{-3}$ \\
48 &  1688.6 &  616.9  &  1077.7 & $-3.6\!\times\!10^{-3}$ \\
52 &  1888.8 &  715.5  &  1182.0 & $-4.6\!\times\!10^{-3}$ \\
\bottomrule
\end{tabular}
\end{table}

We then perform a multichannel CDCC calculation in which $^6$Li is described as an $\alpha$+$d$ cluster, with the breakup continuum discretized into bins up to an $\alpha$-$d$ relative kinetic energy of $12$~MeV and relative orbital angular momentum $\ell\le 2$. The fragment--target optical potentials are taken from Refs.~\cite{alpha2,yyq06}; the surface imaginary part of the deuteron--target potential is removed to avoid double counting with the explicitly treated breakup channels, following Refs.~\cite{jin15-removingsur,jin17-removingsur}. These semi-empirical absorptive inputs are physically motivated but not unique. Table~\ref{tab:cc} presents the resulting decomposition at selected energies. The closure error $|\varepsilon|$ remains below $10^{-2}$~mb ($|\varepsilon|/\sigma_{\rm abs}<10^{-4}$), confirming stable flux bookkeeping in the presence of channel couplings.

\begin{figure}[t]
\centering
\includegraphics[width=\linewidth]{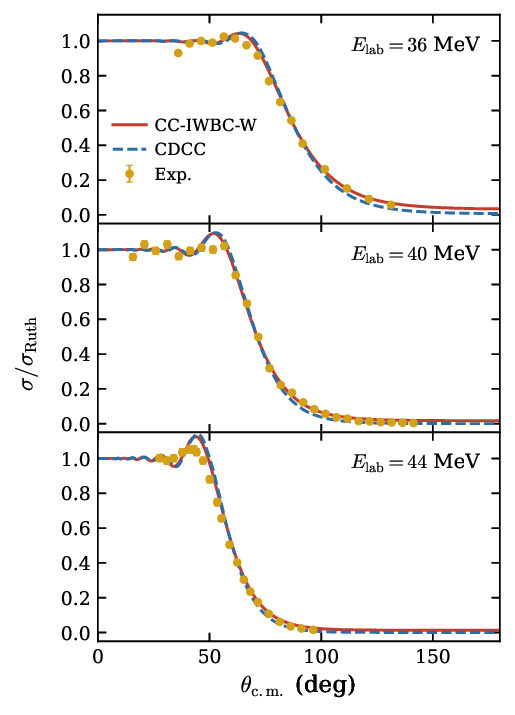}
\caption{Elastic-scattering angular distributions ($\sigma/\sigma_{\rm Ruth}$) for $^6$Li+$^{209}$Bi at 36, 40, and 44~MeV.
Solid line: CC-IWBC-W; dashed line: CDCC; circles: experimental data~\cite{Santra11}.}
\label{fig:elastic}
\end{figure}

Figure~\ref{fig:elastic} compares the coupled-channel elastic-scattering angular distributions with the standard CDCC reference and the experimental data~\cite{Santra11} at 36, 40, and 44~MeV. The present coupled-channel calculation (labeled CC-IWBC-W in the figures; solid line) closely follows the CDCC reference (dashed line) and reproduces the measured angular distributions, confirming that, for the present choice of absorption radius, the additional external imaginary potential does not overdistort the scattering dynamics. Small residual differences suggest that a slightly smaller IWBC radius could further improve the elastic fit, consistent with the expectation that some peripheral processes still occur near the touching configuration. The elastic data are therefore consistent with $r_a=10$~fm as a reasonable reference choice, while the band in Fig.~\ref{fig:fusion} displays the residual sensitivity to nearby values.

\begin{table}[t]
\caption{Multichannel cross-section decomposition for $^6$Li+$^{209}$Bi with $r_a=10$~fm.
$\varepsilon\equiv\sigma_{\rm abs}-\sigma_{\rm fusion}-\sigma_W$.}
\label{tab:cc}
\centering
\begin{tabular}{cccccc}
\toprule
$E_{\rm lab}$ &$\sigma_{\rm R}$ &$\sigma_{\rm abs}$ & $\sigma_{\rm fusion}$ & $\sigma_W$ & $\varepsilon$ \\
(MeV) & (mb) & (mb) & (mb) & (mb) & (mb) \\
\midrule
26 &  61.3 &  39.4 &    1.7 &   37.6 &  $-$0.002 \\
30 & 248.2 & 200.2 &   44.3 &  156.0 &  $-$0.001 \\
34 & 656.3 & 574.0 &  256.4 &  317.6 &  $-$0.002 \\
38 & 1044.8 & 938.3 &  523.6 &  414.7 &  $-$0.007 \\
44 & 1495.0 & 1365.1 &  861.5 &  503.7 &  $-$0.008 \\
48 & 1728.0 & 1586.6 & 1041.9 &  544.7 &  $-$0.005 \\
52 & 1921.6 & 1770.8 & 1193.7 &  577.1 &  \phantom{$-$}0.009 \\
\bottomrule
\end{tabular}
\end{table}

The central result is displayed in Fig.~\ref{fig:decomposition}. Channel couplings produce a marked redistribution of the absorbed flux compared with the single-channel baseline. In that baseline, peripheral absorption dominates across the entire energy range. Once couplings to the $\alpha$+$d$ continuum are included, $f_{\rm fusion}$ increases sharply, from 0.04 at 26~MeV to 0.51 at 36~MeV and 0.67 at 52~MeV, while $f_W$ decreases correspondingly. The two fractions cross near $E_{\rm lab}\approx 36$~MeV, defining a sharp absorption crossover between peripheral loss and inner capture. This crossover, absent in the single-channel baseline, offers a common spatial picture for two long-discussed phenomena of weakly bound complete fusion: at sub-barrier energies, the dramatic rise of $f_{\rm fusion}$ reflects the coupling-induced enhancement of the IWBC-defined inner-capture yield; above the barrier, the persistent peripheral component $\sigma_W$ emerges as a natural contributor to the conventional CF suppression~\cite{Dasgupta04}, as suggested by the comparison with the measured CF data in Fig.~\ref{fig:fusion}. Physically, the continuum couplings modify the effective potential barrier seen by $^6$Li, enhancing the tunneling probability into the inner fusion region. At the same time, the imaginary potentials acting on the fragment--target subsystems continue to remove flux peripherally, so that $\sigma_W$ grows only modestly and its relative share of $\sigma_{\rm abs}$ steadily diminishes.

Comparing Tables~\ref{tab:sc} and~\ref{tab:cc} reveals a second important effect: at above-barrier energies the multichannel $\sigma_{\rm abs}$ is smaller than its single-channel counterpart (e.g., 1770.8 vs 1888.8~mb at 52~MeV), whereas at sub-barrier energies the opposite holds (e.g., 39.4 vs 22.4~mb at 26~MeV). The above-barrier reduction is consistent with the decomposition $\sigma_R=\sigma_{\rm EBU}+\sigma_{\rm abs}$: in the single-channel model, breakup-related flux is absorbed implicitly through $\sigma_W$, whereas in the multichannel calculation the elastic-breakup component is treated explicitly and therefore no longer counted inside $\sigma_{\rm abs}$. Because the multichannel absorptive inputs are semi-empirical and not identical to the single-channel optical potential, this comparison should be read as a physically transparent interpretation rather than a strict channel-by-channel subtraction. At sub-barrier energies, by contrast, the coupling-enhanced barrier penetration more than compensates for this explicit separation of elastic breakup. At the same time, $\sigma_{\rm fusion}$ is markedly enhanced in the multichannel case (1193.7 vs 715.5~mb at 52~MeV). At 52~MeV, for example, $\sigma_W$ drops by about $600$~mb relative to the single-channel baseline, while $\sigma_{\rm fusion}$ increases by about $480$~mb; the remaining difference is naturally associated with flux that is no longer included in $\sigma_{\rm abs}$, most plausibly the explicitly resolved elastic-breakup component.

Figure~\ref{fig:fusion} shows the excitation function of the IWBC-defined inner-capture cross section, which we use as a model proxy for CF. The single-channel result (dashed line) underestimates the measured CF data, as expected in the absence of barrier-distribution effects~\cite{Rowley91,Balantekin98}. The coupled-channel calculation (solid line) yields a pronounced near-barrier enhancement and lies close to the experimental cross sections. The gray band, obtained by varying $r_a$ over 9.5--10.5~fm, confirms that the enhancement is robust against the boundary choice. The reference value $r_a=10$~fm lies slightly above the measured CF data, while $r_a=9.5$~fm gives closer agreement, consistent with the expectation that some peripheral absorption still occurs near the touching configuration. Overall, the IWBC-defined inner-capture cross section reproduces the trend and scale of the measured CF excitation function with only a modest $r_a$ dependence. Combined with the exact identity $\sigma_{\rm abs}=\sigma_{\rm fusion}+\sigma_W$, this agreement gives the peripheral term $\sigma_W$ a concrete interpretation within the adopted model: it represents flux removed peripherally before reaching the inner region, and thus a natural contributor to the suppression of the measured CF yield relative to the inner-capture proxy. At above-barrier energies $\sigma_W$ still carries about one third of $\sigma_{\rm abs}$ ($\sigma_W/\sigma_{\rm abs}\approx 0.33$ at 52~MeV), making it a major contribution to the well-known CF suppression in $^6$Li+$^{209}$Bi.

\section{Summary and outlook}
\label{sec:summary}

We have derived an exact, flux-conserving decomposition $\sigma_{\rm abs}=\sigma_{\rm fusion}+\sigma_W$ for systems described by an IWBC and an external complex potential. Within that model definition, the identity partitions absorption into an IWBC-defined inner-capture term and an external-loss term, and it holds for both single-channel and multichannel systems by virtue of the radial continuity equation.

The application to $^6$Li+$^{209}$Bi shows that channel couplings do not simply increase or decrease the total absorption: they substantially reorganize it.
The inner-capture fraction increases dramatically relative to the single-channel baseline, while the peripheral fraction decreases, leading to a crossover near the Coulomb barrier.
Because $\sigma_{\rm fusion}$, interpreted as an IWBC-defined inner-capture proxy, closely tracks the measured CF excitation function, the peripheral term $\sigma_W$ acquires a concrete physical meaning within the adopted model: it can be interpreted as a major contribution to the conventional CF suppression in $^6$Li+$^{209}$Bi. The sub-barrier enhancement and above-barrier suppression thus emerge as two complementary manifestations of the same coupling-induced spatial redistribution of absorbed flux.

The present formalism provides a diagnostic tool for coupled-channel reaction dynamics: given a CC or CDCC calculation, one can ask not only \textit{how much} flux is absorbed, but \textit{where}.
This spatial partition offers a sharper basis for discussing fusion suppression and enhancement in weakly bound systems~\cite{Canto15}, and provides a natural starting point for connecting phenomenological absorption to dynamically generated effective interactions derived from Feshbach or coupled-channel reduction procedures~\cite{Liu2025ExactDPP,Potel25,KeeleyMackintosh2014}.

Extending the framework to other weakly bound projectiles such as $^7$Li, $^9$Be, and $^{11}$Be will test whether the crossover between peripheral loss and inner capture is a generic feature of complete fusion in these systems.

\section*{Acknowledgements}

This work was supported by the National Natural Science Foundation of China (Grant Nos.\ 12535009 and 12475132), the National Key R\&D Program of China (Contract No.\ 2023YFA1606503), and the Fundamental Research Funds for the Central Universities.

\bibliography{ref}

\end{document}